# Dynamic interface formation in magnetic thin film heterostructures

Shuhei Nakashima, Toshio Miyamachi*, Yasutomi Tatetsu, Yukio Takahashi, Yasumasa Takagi, Yoshihiro Gohda, Toshihiko Yokoyama, and Fumio Komori

Dr. S. Nakashima, Dr. T. Miyamachi, Dr. Y. Takahashi, Prof. F. Komori
The Institute for Solid State Physics, The University of Tokyo
Kashiwa, Chiba 277-8581, Japan
E-mail: toshio.miyamachi@issp.u-tokyo.ac.jp

Dr. Y. Tatetsu, Prof. Y. Gohda
Department of Materials Science and Engineering, Tokyo Institute of Technology,
Yokohama 226-8502, Japan

Dr. Y. Takagi, Prof. T. Yokoyama
Department of Materials Molecular Science, Institute for Molecular Science,
Myodaiji-cho, Okazaki 444-8585, Japan

**Abstract**

Magnetic thin film heterostructures have been widely studied for fundamental interests in the emergence of novel phenomena accompanied by the heterointerface formation as well as their promising practical potential. Combining x-ray magnetic circular dichroism with scanning tunneling microscopy, we show for Mn/Fe thin film heterostructures that the interfacial factors dominating electronic and magnetic properties of the entire system dynamically change with the amount of the Mn overlayer. Element specific magnetization curves of the Fe layer exhibit a two-step spin reorientation transition from out-of-plane to in-plane direction with increasing the Mn coverage. Atomic-scale characterizations of structural and electronic properties in combination with the first-principles calculations successfully unravel the roles of the entangled interfacial factors, and clarify the driving forces of the transition. The first step of the transition at a low Mn coverage is dominantly induced by the formation of FeMn disordered alloy at the heterointerface, and the electronic hybridization with interfacial FeMn ordered alloy is dominant as the origin of the second step of the transition at a high Mn coverage.

## 1. Introduction

In magnetic thin film heterostructures, the heterointerface interaction plays a dominant role for the development of novel electronic and magnetic functionalities, attracting great interest from both fundamental and technological points of view. [1,2] The coupling strength at the heterointerface strongly relies on the interfacial structure on the atomic scale such as atomic roughness, steps and intermixing, which could degrade electronic and magnetic interactions considerably compared to the theoretically predicted ideal interface. [3,4] The impacts of the atomic-scale interfacial structure on the magnetic coupling in magnetic thin film heterostructures were partially discussed in a previous experimental work. [5] Nevertheless, no comprehensive study that takes overall interfacial factors including electronic hybridization on the atomic scale into account, and identifies their individual roles has been conducted so far. To accomplish this purpose, here we use scanning tunneling microscopy (STM) and x-ray absorption spectroscopy/x-ray magnetic circular dichroism (XAS/XMCD) as complementary tools. Successive atomically-resolved STM characterizations of not only structural but also electronic and magnetic properties during the growth of magnetic thin film heterostructures give crucial information on the atomic-scale interfacial factors in the dynamical process of the heterointerface formation. Combined with STM, element-specific, quantitative, and macroscopic observations of electronic and magnetic properties by XAS/XMCD can be linked with microscopic origins of the heterointerface characteristics. The incorporation of structural and more importantly electronic and magnetic properties by microscopic STM observations into macroscopic XAS/XMCD results has been demonstrated for simple systems, e.g., single atoms and molecules on surfaces, [6,7] where several interfacial factors can be negligible, but is still challenging for thin film systems.

Here we focus on fcc Fe thin films grown on Cu(001) with Mn overlayers (Mn/Fe thin film heterosturctures). Electronic and magnetic properties of ferromagnetically-coupled top two layers in the fcc Fe thin film on Cu (001) is quite sensitive to the local lattice strain on the surface even on the atomic scale. [8] Thus, the fcc Fe thin film could highlight the role of atomic-scale interfacial factors when forming the heterointerface with Mn overlayers. Furthermore, collinear and homogeneous exchange coupling at the antiferromagnetic/ferromagnetic Mn/Fe heterointerface is expected likewise the reference system of Mn thin films on the bulk Fe(001) substrate, [9,10] making this system ideal to investigate the micro-macro correlation in magnetic thin film heterostructures. We show from XAS/XMCD measurements that the Fe layer in Mn/Fe thin film heterostructures exhibits a two-step spin reorientation transition (SRT) with the change of the magnetic easy axis from out-of-plane to in-plane direction by the Mn overlayer. Corresponding atomically-resolved STM observations reveal the dynamic change of the dominant interfacial factor for the electronic and magnetic properties of the entire system with the progress of heterointerface formation as the origin of the two-step SRT. The complementary approach by XAS/XMCD and STM successfully disentangles the roles of individual interfacial factors, especially shedding light on the hidden functionality of interface alloying, which has been regarded as the most disruptive effect for the heterointerface, to enhance the magnetic anisotropy in the Fe layer. The results will pave a new way to understand how novel phenomena emerge at the heterointerface on the atomic scale, and to intrinsically improve electronic and magnetic properties of magnetic thin film heterostructure.

## 2. Results and Discussion

**Figure 1a** displays an STM image of 7 ML Fe thin film on Cu(001). The surface morphology is consistent with the previous studies for epitaxial fcc-Fe thin films on Cu(001). [11,12] A high resolution STM image resolves the fcc-Fe(001) lattice with the atomic constant of about 0.26 nm (See the inset of Figure 1a). A similar surface morphology was observed after deposition of 5 ML Mn overlayer, hereafter called Mn(5)/Fe thin film heterostructure, revealing a layer-by-layer growth of Mn on the fcc-Fe surface as shown in Figure 1b. Note that a (12 × 2) surface reconstruction appearing at this Mn overlayer is characteristic of the pure fct Mn surface (See the inset of Figure 1b). [13]

In the XAS/XMCD measurements, the probing depth of the x-ray was firstly checked. Figure 1c displays the XAS of Mn(5)/Fe thin film heterostructure, the thickest sample in the present study, recorded at Mn, Fe and Cu $L_{2,3}$ absorption edges in the NI and GI geometries. We observed the signal from the Cu(001) substrate even in the GI geometry with shorter x-ray probing depth. This ensures that Mn and Fe signals from any Mn/Fe thin film heterostructure are entirely detected. Figure 1d displays reference Fe $L_{2,3}$ XAS and XMCD of a bare 7 ML Fe thin film, i.e., a Mn(0)/Fe thin film heterostructure, recorded at remanence conditions (± 5 → 0 T). The slightly larger remanent XMCD signal in the NI geometry reveals the out-of-plane easy axis of fcc-Fe thin films on Cu(001). [12,14,15] Adding the Mn overlayer drastically changes its XMCD signal. We find for Mn(5)/Fe thin film heterostructure that the remanent XMCD signal in the NI geometry is almost quenched, while the one in the GI geometry keeps its intensity (Figure 1e). This suggests that the Mn overlayer induces the SRT of the underlying Fe layer from out-of-plane to in-plane direction.

To investigate details of the SRT in Mn/Fe thin film heterostructures, we measure magnetization curves of the Fe layer element-specifically with changing the thickness of the Mn overlayer. Figure 2a displays a series of magnetization curves of the Fe layer in Mn(0, 1, 2, 3, 5)/Fe thin film heterostructures recorded in the NI and GI geometries. For Mn(0)/Fe thin film heterostructure, a clear square hysteresis loop is observed only in the NI geometry, and the magnetization saturates in the both geometries at $B = \pm 1.0$ T. This reveals the out-of-plane easy axis of the fcc Fe thin film on Cu(001), in line with the XAS/XMCD results shown in Figure 1d. Its net spin magnetic moment, which dominantly contributes to the amplitude of the magnetization, is quantitatively evaluated to be $0.5 \pm 0.1$ $\mu_B$/atom by applying XMCD sum rules [16,17] to the XAS/XMCD recorded at $B = \pm 5.0$ T. Note that the number of Fe 3d holes, $n_{hole}$, of 3.22 for the sum-rule analysis is evaluated from the comparison of the XAS area with a bulk bcc Fe reference spectrum with $n_{hole}$ = 3.4. [17] Taking magnetically compensated non-collinear antiferromagnetic inner 5 ML in the 7 ML fcc Fe thin films on Cu(001) into account, [12,18,19] the spin magnetic moment of the ferromagnetically-coupled top two layers can be extracted as 1.8 ± 0.3 $\mu_B$/atom. This value is quite close to the one obtained by x-ray resonant magnetic scattering experiments. [18]

Focusing on the impact of the Mn overlayer, we argue that the SRT of the Fe layer from the out-of-plane to in-plane direction takes place in two steps with increasing the Mn coverage. The first change occurs in Mn(1)/Fe thin film heterostructure, i.e., immediately after adding Mn overlayer. For the magnetization curve in the NI geometry, the coercivity drastically decreases (from 0.4 T down to 0.02 T) and the magnetization does not reach to the saturation at B = ± 1 T, even at a maximum magnetic field available in the present study (B = ± 5 T). The magnetization curve in the GI geometry changes its shape, and accordingly a slight decrease in the coercivity was observed (from 0.2 T down to 0.1 T), while keeping the amplitude of the saturation magnetization. The results reveal that 1 ML Mn overlayer significantly lowers the coercivity and saturation magnetization in the NI geometry of the 7 ML Fe thin film.

The second step of the SRT proceeds gradually with further increasing the Mn coverage. In the NI geometry, the coercivity and amplitude of the magnetization at B = ± 1 T gradually decrease up to ~ 3 ML Mn overlayer, and are nearly unchanged thereafter. In contrast, we find that the shape of the magnetization curves in the GI geometry is almost identical in Mn(1, 2, 3, 5)/Fe thin film heterostructures. These changes can be clearly seen in the amplitude of the magnetization at B = ± 1 T as shown in Figure 2b. Since the difference in the amplitude between NI and GI geometries at B = ± 1 T becomes greater up to ~ 3 ML Mn overlayer, the in-plane magnetization is gradually stabilized from 1 to 3 ML Mn overlayer in the second step of the SRT. Indeed, the ratio of the orbital magnetic moment to the spin magnetic moment ($m_{orb}/m_{spin}$) at B = 1 T becomes greater in the GI geometry than the NI one with increasing the thickness of the Mn overlayer (Figure 2c). This suggest that, even though the saturation magnetization is not achieved in the NI geometry at this magnetic field, the out-of-plane orbital magnetic moment is relatively reduced through the SRT process, which in turn results in the enhancement of the in-plane magnetic anisotropy according to the relation between the anisotropy of the orbital magnetic moment and the magnetic anisotropy energy. [20] Note from the nearly constant behavior of the saturation magnetization in the GI geometry shown in Figure 2a and b that the Fe layers in all the Mn/Fe thin film heterostructures contain the spin magnetic moment corresponding to the ferromagnetically-coupled top two layers in the fcc Fe thin films on Cu(001).

We discuss the origin of the first step of the SRT from structural properties of Mn/Fe thin film heterostructures at low Mn coverage. Figure 3a displays a large scale STM image of Mn(0.8)/Fe thin film heterostructure, where the first step just takes places. The heterointerface is at a glance rather flat with wide terraces (> 100 nm), and thus we can exclude atomic steps as the origin of the drastic change in the magnetization curve observed by XAS/XMCD. However, a zoomed STM image shown in Figure 3b reveals that the addition of the 0.8 ML Mn atoms results in considerably rough interface with randomly distributed subnanoscale protrusions. The STM line profile provides the surface roughness of up to ~ 80 pm (Figure 3c), indicating the formation of FeMn disordered alloy during the Mn deposition at room temperature.

The electronic and magnetic properties of fcc Fe thin films on Cu(001) are quite sensitive to the changes in the lattice constant. [7,21] Especially, its out-of-plane magnetic anisotropy emerges due to the expansion of the interlayer spacing between the top two layers relative to the inner layers of 1.78 Å. [22] Since the lattice constant of $Fe_xMn_{1-x}$ alloy is greater than that of intrinsic fcc Fe throughout the composition x (from 3.605 to 3.796 Å), [23] the alloy with a greater in-plane lattice constant can expand the in-plane lattice of the interface Fe layer. This would shorten the interlayer spacing between the top two Fe layers, and accordingly reduce its out-of-plane magnetic anisotropy. Furthermore, the heterointerface roughness broadens the distributions of the Fe in-plane lattice constant and the interlayer spacing, which hinders a long range order of the ferromagnetic Fe layers. The inhomogeneity could lower the coercivity of the Mn(1)/Fe thin film heterostructure in the NI geometry as observed in Figure 2a. Thus, the origin of the first step of the SRT is reasonably attributed to the lattice change and roughness caused by the formation of the surface disordered alloy.

To understand the origin of the second step of the SRT, we first investigate the structural properties of Mn/Fe thin film heterostructures in the transition region from 1 to 3 ML Mn overlayers. Figure 4a displays STM images of Mn/Fe thin film heterostructures with 1.0, 1.5 and 1.8 ML Mn overlayer. The formation of the heterointerface proceeds in a layer-by-layer fashion, i.e., an additional second layer (level-2) covers over the first one (level-1) with increasing the Mn coverage. Corresponding high-resolution STM images on level-1 are displayed in Figure 4b, c, and d. In contrast to the heterointerface composed of only the disordered alloy below 1.0 ML Mn coverage (Figure 3a and b), the ordered alloys with the (4 × 2) reconstruction, and minutely with the (4 × 4) one (< 1.5 ML Mn overlayer) start to appear. The formation of the ordered alloy is also confirmed on level-2 as shown in a high-resolution STM image of Mn(1.8)/Fe thin film heterostructure (Figure 4e). Since Mn-based surface alloys can generally exhibit a variety of ordered structures such as, $c(2 \times 2)$, $c(8 \times 2)$, $c(12 \times 8)$ and $p2mg(4 \times 2)$ reconstructions with different Mn compositions,[24] the structural transition from disordered alloy to ordered one with the increase in the Mn coverage could reflect the increased Mn composition in the ordered alloy. Actually, no Mn XMCD signal was confirmed for all Mn/Fe thin film heterostructures in the present study possibly due to a noncolinear antiferromagnetic order in the FeMn alloys,[25, 26] ensuring the high composition of Mn not only in the ordered alloy but also in the disordered alloy.[27,28] Note that the magnetization in the alloy interface layer can be parallel to those in the underlying Fe layers as in the case of the interface between a ferromagnetic Co and an FeMn alloy.[29]

The atomic resolution capability of STM in real space further allows us to extract important aspects of the heterointerface formation. Figure 4f displays fractions of the ordered alloy in level-1 and level-2 as a function of the Mn coverage. While the only disordered alloy exists at the Mn overlayer less than 1 ML, the fraction of the ordered region in level-1 monotonously increases once the Mn overlayer exceeds 1 ML (See also Figure 4b, c, and d). The ordered region in level-2 also increases with increasing the Mn coverage, and its fraction is greater than that in level-1. Extrapolating the fractions of the ordered region gives the full coverage of the ordered alloy in both level-1 and level-2, i.e., the completion of the heterointerface, at ~ 3 ML Mn overlayer. Accordingly, pure Mn layers with a characteristic $(12 \times 2)$ reconstruction as shown in Figure 1b start to appear from the third layer. These results reveal that the Mn rich ordered alloy promotes the dynamic formation of the heterointerface. More importantly, the agreement of the threshold Mn thickness for completing the interface ordered alloy with that for establishing the second step of the SRT (~ 3 ML) suggests a strong correlation between the formation of the ordered alloy and the stabilization of the in-plane magnetic anisotropy of the Fe layer in Mn/Fe thin film heterostructures.

The observed rapid drop of the Fe coercivity at the first step of the SRT could be caused by both the reduction of interlayer distance between the interface two Fe layers and the significant structural roughness of the disordered alloy. The second step gradually enhances the in-plane magnetic anisotropy of the Fe layer in accordance with the gradual ordered-alloy formation. This slow change indicates that the structural corrugation of the ordered alloy itself does not play a dominant role in the second step. We here relate the interface 3d hybridization between the ordered alloy and the Fe layer near the Fermi level to the establishment of the second step of the SRT. For this purpose, we investigated the *dI/dV* spectra of the Mn/Fe thin film heterostructures with 0, 1, 3, and 5 Mn overlayers as shown in Figure 5. Indeed, the *dI/dV* spectra of Mn(0, 3, 5)/Fe thin film heterostructures exhibit several electronic states near the Fermi level, especially the peak structure commonly appears at ~ − 0.2 eV. Note that the peak structure at ~ − 0.2 eV of Mn(5)/Fe thin film heterostructure, i.e., the pure fct Mn surface, would originates from out-of-plane-oriented $3d_{3z^2-r^2}$ orbital. [8,30] Accordingly the peak of Mn(3)/Fe thin film heterostructure, the ordered FeMn alloy, at ~ − 0.2 eV is expected to have the same origin. The absence of the clear electronic states in the *dI/dV* spectrum of Mn(1)/Fe thin film heterostructure near the Fermi level is reasonable since the roughness of the disordered alloy significantly widens the electronic states characteristic of the ordered alloy. With gradually increasing the Mn coverage and thus the ordering of the FeMn alloy at the heterointerface, the peak structure at ~ − 0.2 eV in the *dI/dV* spectrum becomes significant.

For the fcc Fe thin film, our first-principles calculations reveal that the origin of the peak in the local density of states (LDOS) at ~ − 0.2 eV is the $3d_{3z^2-r^2}$ minority spin state (Figure 6). Since the saturation magnetization of the Fe layer in the GI geometry is nearly constant as shown in Figure 2b, the hybridization with the ordered or disordered FeMn alloys at the heterointerface do not largely change the overall Fe 3d electronic states determining the magnetization across the SRT. This is supported by the observation that there is no significant change in the spectral shape of Fe L$_3$ XAS among Mn/Fe thin films heterostructures (Figure S1 in Supporting Information). The observed peak at ~ − 0.2 eV in the Fe *dI/dV* spectrum would be then preserved irrespective of the Mn overlayer. This leads to the efficient hybridization between the Fe layer and Mn overlayer via the out-of-plane orbitals commonly located around this energy. On the other hand, the in-plane $3d_{3z^2-r^2}$ orbital, which dominantly contributes to the out-of-plane magnetic anisotropy of the fcc Fe film [31] due to a larger spin-splitting than another in-plane $3d_{xy}$ orbital, has no clear LDOS near the Fermi energy. Thus, this orbital does not hybridize with the Mn overlayer at ~ − 0.2 eV.

At the heterointerface, the electronic state of the FeMn ordered alloy locating at ~ − 0.2 eV as seen in the *dI/dV* spectrum of Mn(3)/Fe thin film heterostructure can hybridize with the Fe 3d orbitals of the Fe layer at the same energy, and accumulate their electrons to this energy. In addition to the out-of-plane $3d_{3z^2-r^2}$ orbital, such an electron accumulation can happen for the minority-spin states with significant LDOS near the Fermi energy, i.e., $3d_{yz}$ and $3d_{zx}$ with the out-of-plane orbitals and $3d_{xy}$ with the in-plane orbital according to the Fe orbital- and-spin-resolved LDOS (see Figure 6). Thus, we can expect the dominant accumulation of electrons in the out-of-plane 3d orbitals of the Fe layer from the energy range between ~ − 0.2 eV and the Fermi energy. This would enhance the in-plane magnetic anisotropy as a consequence of the increase in the LDOS of the out-of-plane Fe 3d orbitals near the Fermi energy. [31-34] Furthermore, as shown in Figure 6, this tendency is valid even if the thickness of the Fe layer changes (6, 7, and 8 ML), e.g., due to the alloy formation.

## 3. Conclusion

In summary, we investigated structural, electronic and magnetic property of Mn/Fe thin film heterostructures. Fully taking the advantage of element-specific and quantitative magnetic characterization capabilities of XAS/XMCD, we find that the Fe layer in Mn/Fe thin film heterostructure exhibits a two-step SRT from out-of-plane to in-plane magnetization with increasing the Mn coverage. The origin of the observed two-step SRT is identified by separately evaluating the roles of entangled interfacial factors using STM with atomic-resolution imaging and spectroscopic capabilities. At low Mn coverages (< 1 ML), a considerably rough heterointerface due to the formation of the disordered alloy drastically weakens the out-of-plane magnetization of the Fe layer, accordingly triggering the first step of the SRT. In the second-step SRT, the in-plane magnetic anisotropy of Mn/Fe thin film heterostructures is gradually enhanced up to ~ 3 ML Mn coverage. With the help of the first-principles calculations, we attribute the stabilization of the in-plane magnetization dominantly to the electronic hybridization of the Fe layer with the ordered alloy at the heterointerface. The results demonstrate how effectively microscopic characterizations by STM can be integrated into macroscopic ones by XAS/XMCD to achieve a comprehensive understanding of the dynamic heterointerface formation. Furthermore, a considerable enhancement of the magnetic anisotropy of the Fe layer across the second step SRT will provide a new perspective on the materials design including the interfacial alloy to reinforce magnetic thin film heterostructures.

## 4. Experimental Section

All the samples were prepared in an ultrahigh vacuum (UHV, $< 2.0 \times 10^{-10}$ Torr). A clean Cu(001) surface was prepared by cycles of Ar$^+$ sputtering and annealing at 720 K. Thin film heterostructures of Mn and Fe were fabricated at room temperature on Cu(001) by molecular beam epitaxy from high-purity Fe (99.998 %) rods and Mn (99.99 %) pieces. The thickness of the Fe layer was fixed to ~ 7 ML and that of the subsequently deposited Mn overlayer ranges from 0 to 5 ML. Here, ML is defined as the Cu(001) surface atomic density. The XAS/XMCD measurements were carried out at BL 4B of UVSOR-III [35] in a total electron yield mode at 80 K. External magnetic fields B up to ± 5 T were applied parallel to the incident x-ray. The circular polarization of the incident x-ray was set to 65 %. The XAS were recorded in the normal (NI: $\theta = 0°$) and the grazing (GI: $\theta = 55°$) incident geometries. Note that $\theta$ is the angle between the sample normal and the incident x-ray. The XMCD is defined as $I_+ - I_-$, where $I_+$ and $I_-$ denote the XAS intensity at the Fe and Mn $L_{2,3}$ adsorption edges with the photon helicity parallel and antiparallel to the external magnetic field. The STM measurements were performed at 80 K. The tips were prepared by electrochemical etching of W wires and cleaned by subsequent flashing in UHV. Topographic images were obtained in a constant current mode and the differential conductance spectra of the tunneling current, *dI/dV*, were recorded using a lock-in technique with a bias-voltage modulation of 20 mV and 733 Hz.

First-principles calculations for spin- and orbital- resolved LDOS of 6, 7 and 8 ML fcc Fe thin films on Cu(001) were performed using a computational code OpenMX. [36,37] Our model structure is composed of 6, 7, and 8 Fe layers on the substrate of 7 Cu layers in the unit cell. We set a 10 Å vacuum region on the Fe layers. The lattice parameter of the fcc Cu substrate is set to 1.8195 Å determined by DFT. Note that the calculated lattice parameter of fcc Fe, 3.644 Å, is close to the calculated lattice parameter of fcc Cu. We used a $60 \times 60 \times 1$ *k*-point mesh in the first Brillouin zone, and 500 Ry cutoff energy for solving the Poisson equation. The convergence criteria for the maximum force on each atom and the total energy were $10^{-4}$ hartree/bohr and $10^{-7}$ hartree, respectively. The atomic positions of Fe and Cu atoms were fully relaxed except for the Cu atoms in the lower 5 layers. The top two Fe layers couple ferromagnetically, and the second top layer shows an antiferromagnetic coupling with the third top layer, which is consistent with the previous theoretical and experimental results in this system. [19, 22]

**Supporting Information**

Supporting Information is available from the Wiley Online Library or from the author.

**Acknowledgements**

This work was partly supported by JSPS KAKENHI for Young Scientists (A), Grant No. 16H05963, for Scientific Research (B), Grant No. 26287061, the Hoso Bunka Foundation, Shimadzu Science Foundation, Iketani Science and Technology Foundation, Nanotechnology Platform Program (Molecule and Material Synthesis) of the Ministry of Education, Culture, Sports, Science and Technology (MEXT), and the Elements Strategy Initiative Center for Magnetic Materials (ESICMM) under the outsourcing project of MEXT. S. N. and Y. Takahashi were supported by the Grant-in-Aid for JSPS Research Fellow.

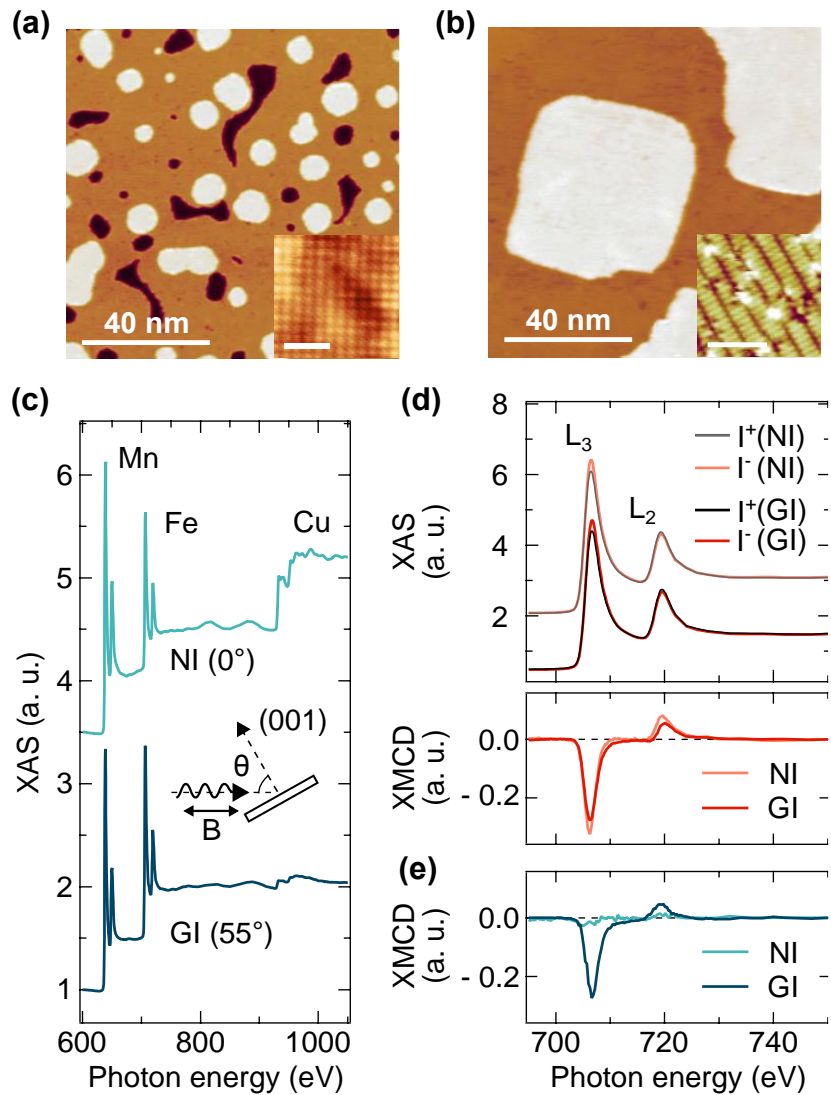

**Figure 1.** STM images of (a) 7 ML Fe thin film on Cu(001) and (b) Mn(5)/Fe thin film heterostructure. The insets display (a) fcc-Fe(001) lattice and (b) Mn(12 × 2) reconstructed surfaces with scale bars of 1 and 5 nm, respectively. (c) Mn, Fe and Cu $L_{2,3}$ XAS of Mn(5)/Fe thin film heterostructure recorded in the NI and GI geometries. (d) Fe $L_{2,3}$ remanent XAS and XMCD of Mn(0)/Fe thin film heterostructure and (e) Fe $L_{2,3}$ remanent XMCD of Mn(5)/Fe thin film heterostructure recorded in the NI and GI geometries.

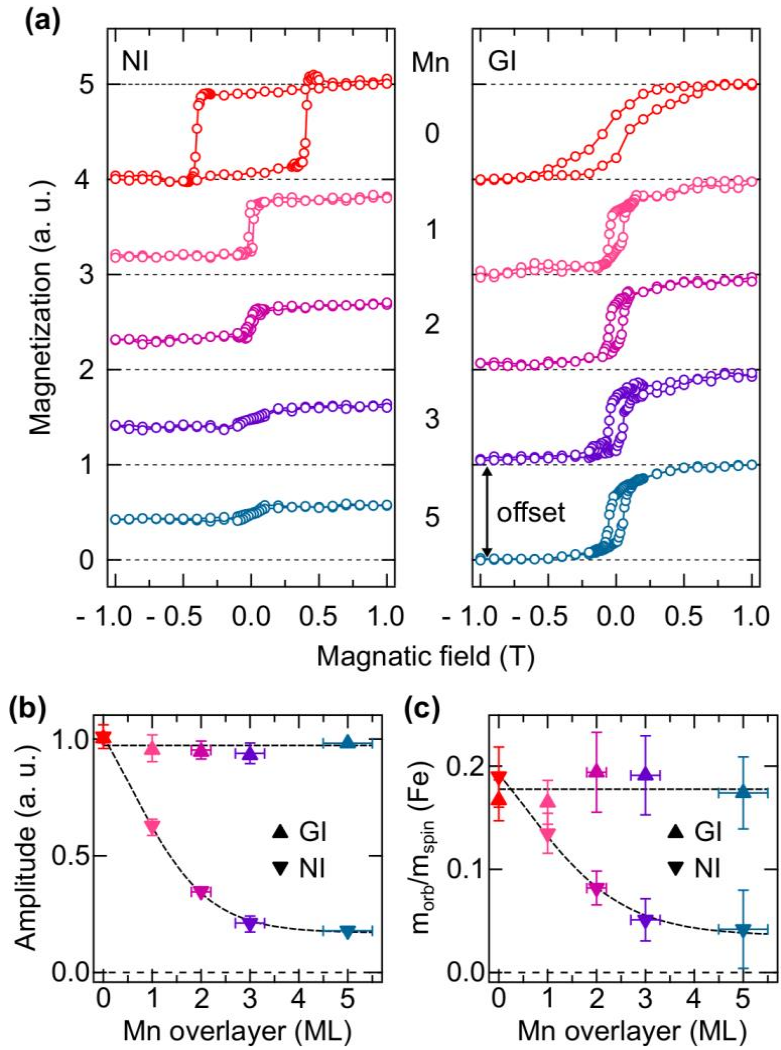

**Figure 2.** (a) Magnetization curves of the Fe layer in Mn/Fe thin film heterostructures with 0, 1, 2, 3, and 5 ML Mn overlayers (from top to bottom) measured in the NI and GI geometries. The $L_3$ XAS peak intensity normalized by the $L_2$ one is plotted one as a function of B. In the GI geometry, the magnetic field is applied along the [100] direction of the Cu(001) substrate. (b) The amplitude of the magnetization between $B = \pm 1$ T and (c) the value of $m_{orb}/m_{spin}$ at $B = 1$ T as a function of the Mn overlayer in the NI and GI geometries. The error bar is smaller than the symbol size if not seen. The horizontal error bars represent the deviations of the Mn thickness estimated by the quartz-crystal oscillator and the Mn L-edge jump. The dashed lines are guides to the eye.

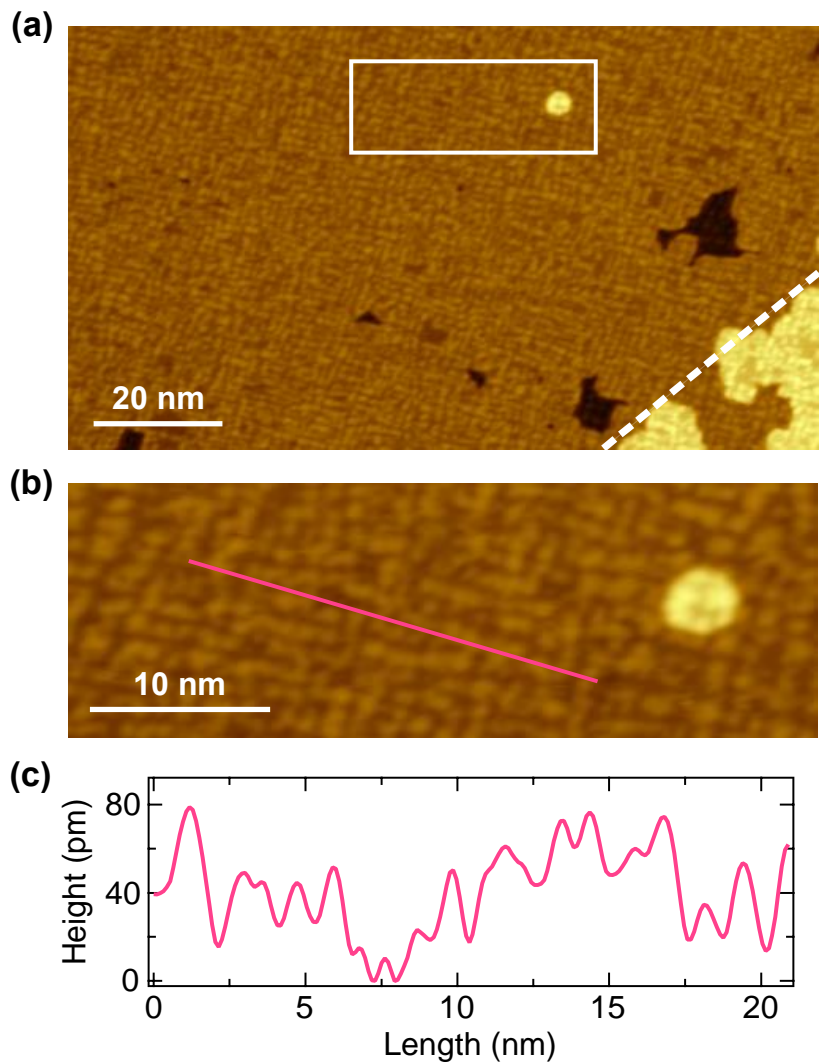

**Figure 3.** (a) STM image of Mn(0.8)/Fe thin film heterostructure. The dashed line represents the atomic step. (b) Zoomed STM image of the rectangle in (a). (c) STM height profile along the pink line in (b).

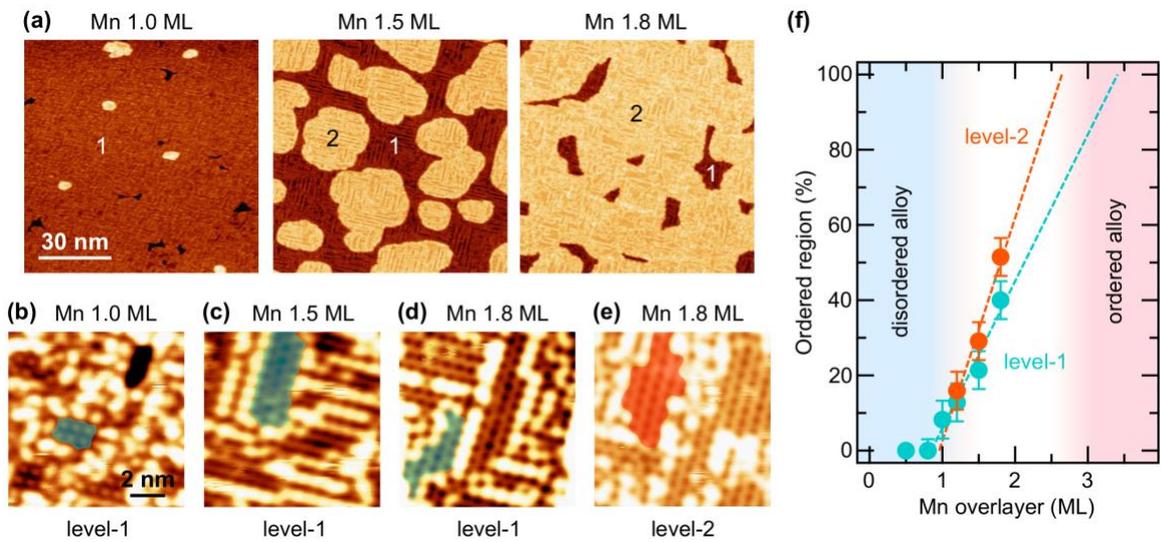

**Figure 4.** (a) STM images of Mn(1.0, 1.5, 1.8)/Fe thin film heterostructures. High-resolution STM images of Mn/Fe thin film heterostructures with (b) 1.0, (c) 1.5, (d) 1.8 ML Mn overlayer in level-1 and (e) 1.8 ML Mn overlayer in level-2, respectively. The ordered regions are partially colored by blue (level-1) and red (level-2). The disorder alloys can be seen in each image as bright protrusions. (f) Statistical plots of the fraction of the ordered alloy with the (4 × 2) and minutely (4 × 4) reconstructions in level-1 and level-2 as a function of the Mn overlayer. The dashed lines are linear extrapolations of the fractions of the ordered alloy.

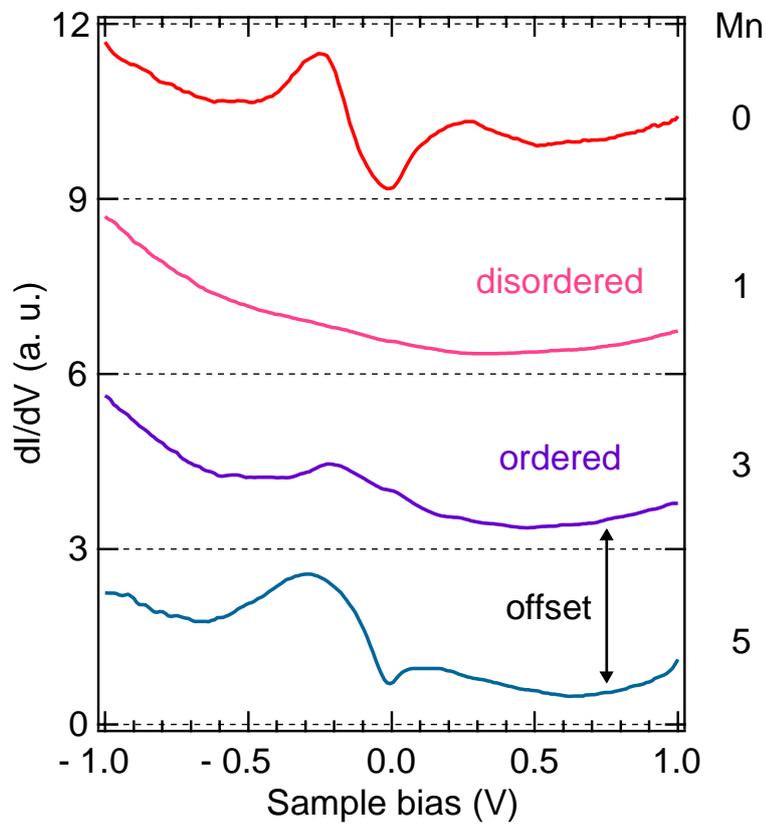

**Figure 5.** A series of the dI/dV spectra recorded on Mn(0, 1, 3, 5)/Fe thin film heterostructures (from top to bottom). The surfaces of the Mn/Fe thin film heterostructures with 1, 3 and 5 ML Mn overlayers are dominantly composed of the disordered alloy, ordered alloys, and pure fct Mn, respectively.

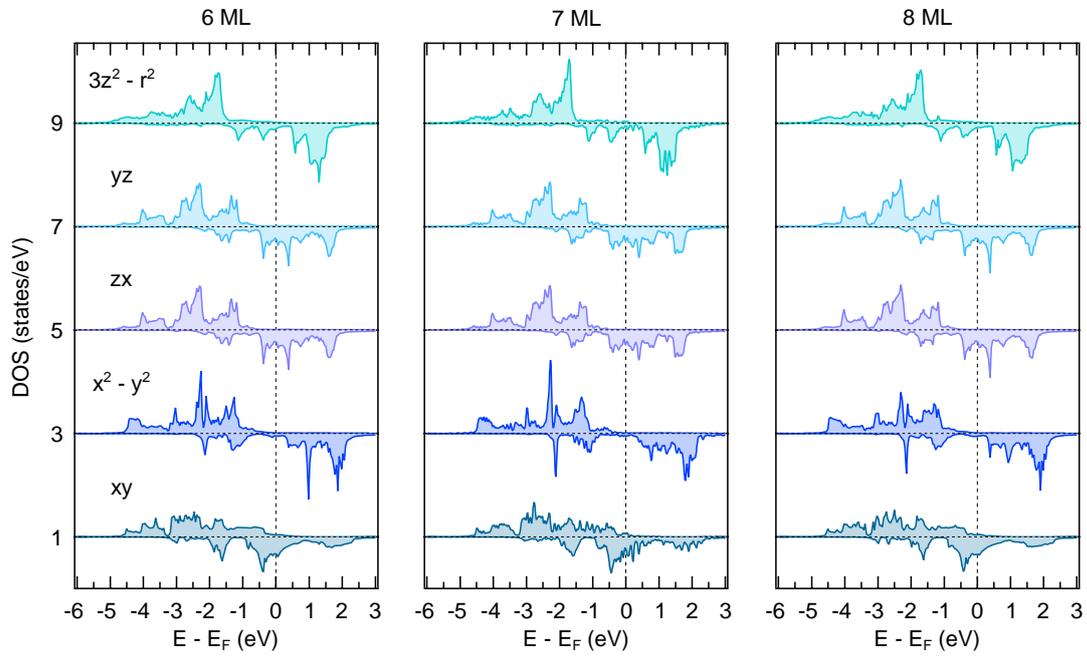

**Figure 6.** Spin- and -orbital resolved LDOS of 6, 7, and 8 ML fcc Fe thin films on Cu(001).

Supporting Information

**Dynamic interface formation in magnetic thin film heterostructures**

Shuhei Nakashima, Toshio Miyamachi*, Yasutomi Tatetsu, Yukio Takahashi, Yasumasa Takagi, Yoshihiro Gohda, Toshihiko Yokoyama, and Fumio Komori

Dr. S. Nakashima, Dr. T. Miyamachi, Dr. Y. Takahashi, Prof. Komori
The Institute for Solid State Physics, The University of Tokyo
Kashiwa, Chiba 277-8581, Japan
E-mail: toshio.miyamachi@issp.u-tokyo.ac.jp

Dr. Y. Tatetsu, Prof. Y. Gohda
Department of Materials Science and Engineering, Tokyo Institute of Technology,
Yokohama 226-8502, Japan

Dr. Y. Takagi, Prof. T. Yokoyama
Department of Materials Molecular Science, Institute for Molecular Science,
Myodaiji-cho, Okazaki 444-8585, Japan

**Spectral shape of Fe L$_3$ XAS of Mn/Fe thin film heterostructure**

To investigate the impact of the Mn overlayer on the electronic properties of the underlying Fe layer in Mn/Fe thin film heterostructure, we check the spectral shape of Fe L$_3$ XAS with different Mn overlayer thicknesses. The XAS at Fe L$_{2,3}$ absorption edges (2p → 3d) reflects the local density of states (LDOS) of Fe 3d orbitals near the Fermi energy especially in the unoccupied bands. In case that the Mn overlayer affects the electronic properties of the Fe layer significantly, electronic modifications in the Fe 3d states can be recognized in the spectral shape of the Fe L$_{2,3}$ XAS as a broadening of the spectral width, or as an emergence of additional shoulder peak structures with increasing Mn overlayer.[1,2] Figure S1a and b display Fe L$_3$ XAS of Mn(0, 1, 2, 3, 5)/Fe thin film heterostructures recorded in the NI and GI geometries. We find no clear coverage dependence of the Mn overlayer for the spectral shape of Fe L$_3$ XAS. Thus, we can safely conclude that adding Mn overlayer does not essentially change the distribution of the LDOS of Fe 3d orbitals in the Fe layer near the Fermi energy. Note from the antiferromagnetic order in the ordered and disordered alloys that the high composition of Mn is expected for both of them. [3,4] Thus, the Fe atoms in the heterointerface would be minute and not affect the spectral shape of Fe L$_3$ XAS.

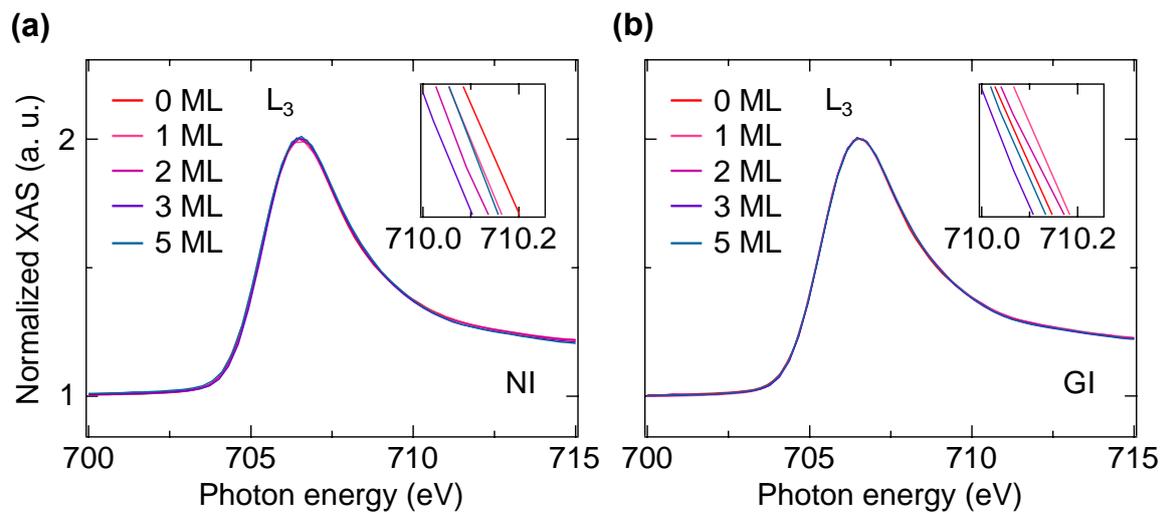

**Figure S1.** Fe $L_3$ XAS of Mn(0, 1, 2, 3, 5)/Fe thin film heterostructures in the (a) NI and (b) GI geometries. The spectra are normalized by the $L_3$ peak intensity located at ~706 eV.